\documentclass[aps,pre,twocolumn,showpacs,amssymb]{revtex4}

\usepackage[dvips]{graphicx}
\usepackage{dcolumn}
\usepackage{bm}

\def\gs{\gtrsim}
\def\ls{\lesssim}
\def\be{\begin{equation}}
\def\en{\end{equation}}                  
\def\p{\partial} 

\newcommand{\av}[1]{\langle{#1}\rangle}\def\bea{\begin{eqnarray}}
\def\ena{\end{eqnarray}}
\newcommand{\ppp}[3]{{\bigg(}\frac{\partial {#1}}{\partial {#2}}{\bigg )}_{#3}}
\def\ge{> \kern -12pt \lower 5pt \hbox{$\displaystyle =$}}
\def\le{< \kern -12pt \lower 5pt \hbox{$\displaystyle =$}}
\def\gs{> \kern -12pt \lower 5pt \hbox{$\displaystyle{\sim}$}}
\def\ls{< \kern -12pt \lower 5pt \hbox{$\displaystyle{\sim}$}}

\begin{document}
\title{High-speed observation of 
the piston effect near the gas-liquid critical point}
\author{Yuichi Miura$^1$, Mitsuru Ohnishi$^2$, 
Shoichi Yoshihara$^2$, 
Katsuya Honda$^3$, 
Masahiro Matsumoto$^3$, 
Jun Kawai$^3$, Masamichi Ishikawa$^4$, 
Hiroto Kobayashi$^5$, and Akira Onuki$^6$}
\affiliation{
$^1$Department of Physics, Nagoya University, Nagoya 464-8602, Japan\\
$^2$Japan Aerospace Exploration Agency, Chofu, Tokyo 182-8522, Japan\\
$^3$Mitsubishi Research Institute Inc., Chiyoda-ku, Tokyo 100-8141, Japan\\
$^4$Interdisciplinary Graduate School of Science and Engineering, 
Tokyo Institute of Technology, Yokohama 226-0026, Japan\\  
$^5$College of Engineering, Chubu University, Kasugai 487-8501, Japan\\@
$^6$Department of Physics, Kyoto University, Kyoto 606-8502, Japan}
\begin{abstract}
We measured  high-speed sound propagation in a 
near-critical fluid 
using a ultra-sensitive interferometer  to 
investigate  adiabatic changes of fluids 
on acoustic timescales.  A sound emitted 
by very weak continuous heating 
caused   a stepwise adiabatic change 
at its front with a density change 
of order $10^{-7}$g$/$cm$^3$ and 
a temperature change of order  $10^{-5}$deg. 
Very small heat  inputs 
at a heater produced  short acoustic pulses with width 
of order $10\mu$sec, which were   broadened 
as they moved through  the cell and encountered  
with the boundaries. The  pulse broadening became
enhanced  near the critical point. 
We also  examined  theoretically 
how sounds are emitted from a heater 
and how applied heat is transformed into 
mechanical work. Our predictions 
well agree with our data.
\end{abstract}
\pacs{05.70. Jk, 64.70. Fx, 62.60. +v, 65.40. De}
\maketitle

 Thermal equilibration 
in one-component fluids takes place increasingly faster near 
the gas-liquid critical point  at fixed volume  
\cite{
Straub,Ferrell,Hao,Gammon,Beysens,Meyer,Moldover,Carles,Maekawa,Onukibook}, 
despite the fact that the thermal 
diffusion constant $D$ tends to zero  at the criticality. 
This is because  the thermal diffusion layer at the 
 boundary expands 
and   sounds emitted     cause 
   adiabatic compression and heating 
in the whole cell after many traversals in the  container.
 This  heating mechanism 
is much intensified near the critical point 
due to the critical enhancement of 
thermal expansion of the layer. 
If the boundary temperature $T_w$ 
is fixed,  the interior 
temperature approaches $T_w$ 
on  the timescale of the piston time \cite{Ferrell}, 
\be 
t_1= L^2/4(\gamma-1)^2D, 
\en 
where $L$ is the 
cell length  and $\gamma=C_p/C_V$ 
is the specific-heat ratio growing near the critical point. 
This time is  much shorter  than 
the isobaric equilibration time $L^2/4D$ by the very small factor 
  $(\gamma-1)^{-2}$ \cite{ratio}.

The previous  experiments  have  detected only 
 slow  temperature and density changes in the interior region 
on timescales  of order 1 sec. 
The aim of this letter is to report  ultra-sensitive,  
high-speed  observation of sound 
propagation through a cell filled with  CO$_2$  
on the critical isochore close 
to the critical point 
 $T_c=304.12$K. We can detect density 
changes of order $10^{-8}$g$/$cm$^3$ 
taking place on a timescale of $1  \mu$sec. 
In confined fluids, adiabatic changes are  usually 
 caused   by   a mechanical piston, 
but they can also be achieved by a heat input 
through a boundary 
\cite{Straub,Ferrell,Hao,Gammon,Beysens,Meyer,Moldover}. 
The second thermo-acoustic 
method is particularly efficient near the critical point. 
These adiabatic  processes 
are very fundamental  and ubiquitous, but 
we are not aware of any  experiments 
detecting   the  underlying fast  acoustic processes.

\begin{figure}[h]
\includegraphics[scale=0.38]{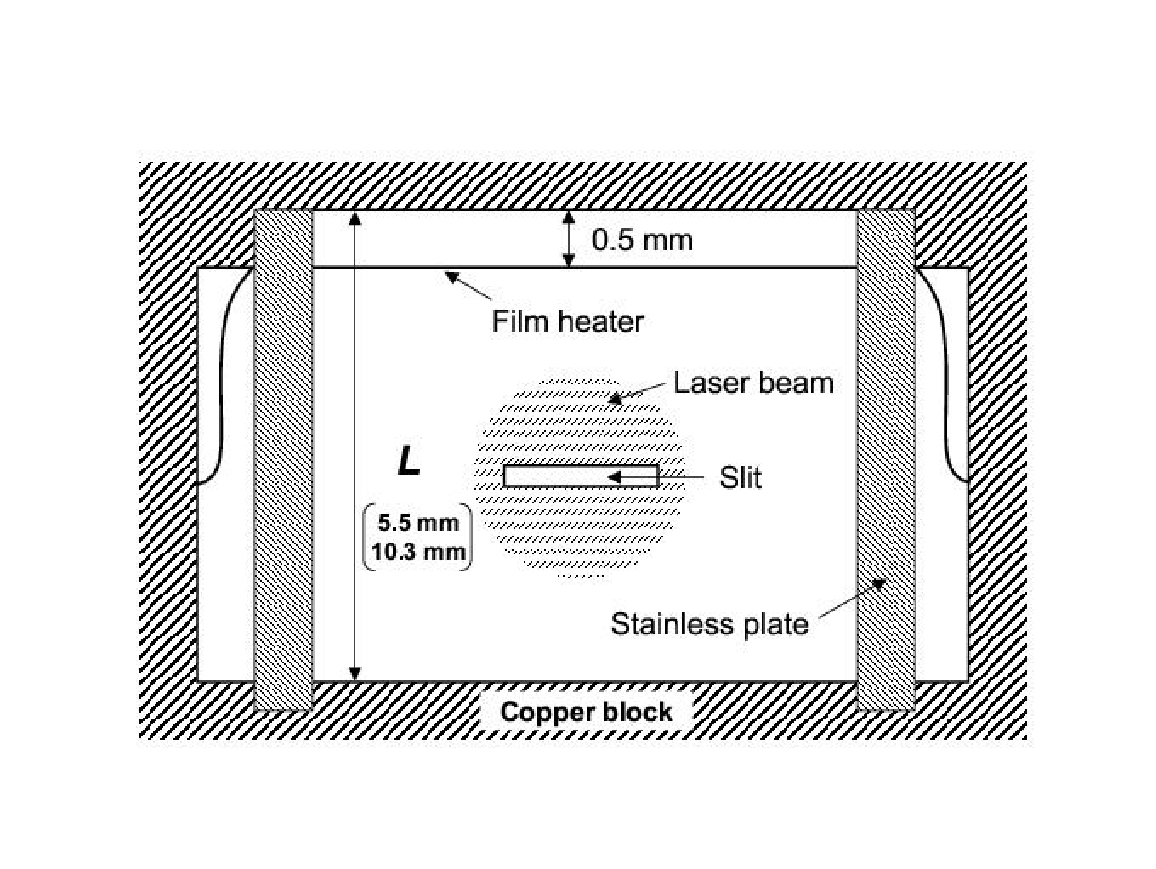}
\caption{
Front view of experimental set-up. 
Cell length $L$ is either $10.3$mm  or $5.5$mm. 
A dc current is sent to a film heater placed $0.5$mm 
below the top boundary. 
Light passing through the slit of width 0.25mm 
is used to detect 
small density changes using interferometry. Gravity is downward.}
\end{figure}

Our experimental  set-up 
is displayed  in Fig.1, where the fluid 
temperature was controlled  with precision of 
$\pm1$mK using a Pt-thermometer 
and a four-wire resistance bridge \cite{exp}.  
The upper and lower plates  have area 1 cm$^2$ 
and are made of Cu with 
high thermal conductivity $\lambda_{\rm Cu}/k_B = 
2.8\times 10^{23}/$cm  sec. The cell length $L$ 
was either  $10.3$mm  or $5.5$mm. 
A thin NiCr-foil  heater with thickness $3\mu$m  
was  placed  $d=0.$5mm below the top plate. 
The  heat capacity of the foil  is 
so small ($C_h/k_B = 8.4\times 10^{19}/$cm$^2$ per unit area) 
and  the  generated heat   
was  almost released to the fluid \cite{Ch}.   
The side walls are made of 
stainless steel,  whose thermal conductivity 
is $3.9\%$ of that of Cu.  We thus neglect 
heat flow to the side walls.     A  laser beam was sent at  
the cell center and a  Twymann-Green interferometer was used to 
detect small density changes  in a slit region 
of width 0.25mm. Individual  signals 
were  very noisy and  the data points 
in the following figures are the averages over 418 heat pulses 
successively generated  in  2 seconds. 
Using sound pulses we first  measured the sound speed 
$c$. For  both  $L= 10.3$mm and $5.5$mm, 
our data on $c$ closely agree  with previous ones 
 in the  range $T/T_c-1\gs 10^{-4}$ \cite{sound},  
but  become independent of 
$T-T_c$ closer to the critical point 
 because of the frequency-effect 
(inherent to  short-time pulses) 
and/or the gravity effect \cite{Hohenberg}.

\begin{figure}[h]
\includegraphics[scale=0.45]{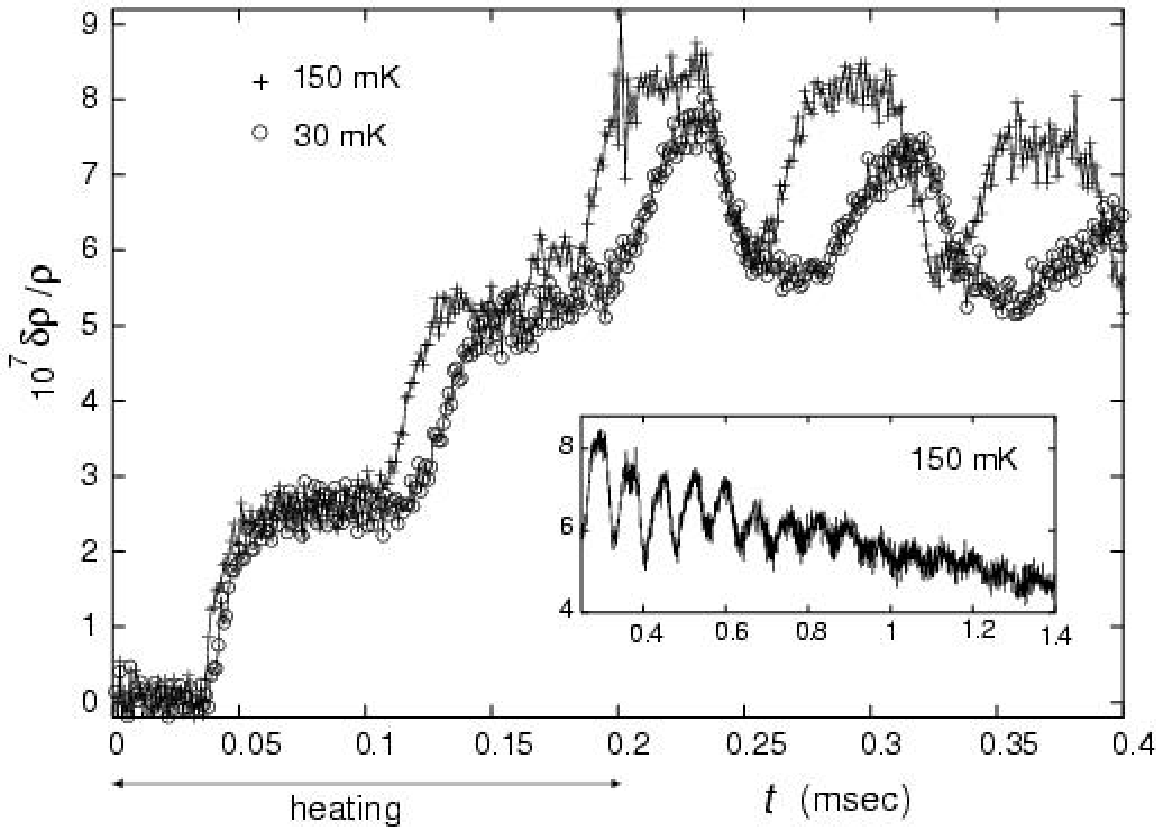}
\caption{
Normalized density change $\delta\rho(t)/\rho$ 
at the cell center in the time region 
$0<t<0.4$msec for $T-T_c=150$mK 
and $30$mK,   produced by 
continuous heating in  $0<t<0.2$msec in a cell of 
$L=1.03$cm.   Inset: Long-time 
behavior for $T-T_c=150$mK in the time region 
$0.2<t<1.4$msec.}
\end{figure}
\begin{figure}[h]
\includegraphics[scale=0.5]{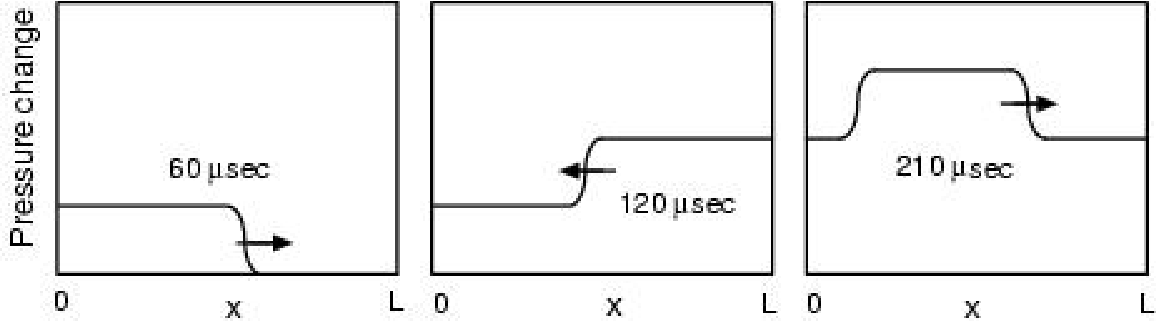}
\caption{
Profiles of pressure deviations 
propagating at the sound speed 
at  $T-T_c=150$mK.  Heating is stopped at $t=200\mu$sec, 
which results in a trapezoidal pulse (right).  
}
\end{figure}

In Fig.2 we show acoustic  
density variations $\delta\rho(t)$  at 
the cell center $x=L/2$  in a cell of $L=1.03$cm  at  
$T-T_c=150$ and $30$mK, where 
$c\cong 140$ and $123$m$/$sec, respectively.  
Continuous heating was  applied in the time 
region $0<t<0.2$msec and a sound signal 
arrived   at the cell center only 
for $t>(L/2-d)/c \sim 40\mu$sec. 
 The total heat supplied was $Q=367$erg.
The inset of Fig.2 displays a rapid  
relaxation at $T-T_c=150$mK after switching-off  the heater.  
Fig.3 illustrates    spatial  profiles of the 
 pressure deviation  $\delta p$, whose 
inhomogeneity remains small even in the 
thermal diffusion layers. 
Notice that $\delta \rho$ takes 
 large negative values in the layers. 
The thickness of the layers was 
only $2.6\times 10^{-5}$cm at $t=0.2$msec 
 for $T-T_c=150$mK. 
Right after switching-on 
the heater,  step-like sounds 
were  emitted on  both sides  and the one moving upward was   
reflected at the top  at $t\cong d/c$. The 
two pulses  merged     
for $t>2d/c \sim 10\mu$sec. 
After  sweeping 
of the sound,   we can see a stepwise increase in 
the density  at $t \cong L/2c,3L/2c$, and $5L/2c$. 
After the switching-off at 
$t=0.2$msec,   a trapezoidal pulse appeared  moving  
back and forth in the cell.  
The  trapezoidal part should be diminished if the heating 
time would be a multiple of $2L/c$. 
The interior temperature was  slightly increased by  
 $\delta T=(\p T/\p \rho)_s\delta \rho$ 
and  a  heat flow  was  induced 
through the boundaries.  In the present case $ 
\delta T \sim 0.2 T \delta\rho/\rho \sim 10^{-5}$deg.  
As can be seen 
in the inset  of Fig.2,  the 
subsequent relaxation in  the interior  region 
was  very rapid. 
It is  the same as  that after 
 cooling of the boundaries  
on long times scales 
($t \gg L/a)$, so   it is  
an  adiabatic process caused by   
 contraction of the thermal 
diffusion layers.

In Fig.4 we show  $\delta\rho(t)$  
 after very short heating in a cell 
of $L=5.5$mm  at $T-T_c=500$ and $100$mK, where 
$c\cong 149$ and $133$m$/$sec, 
respectively. A dc electric current 
of duration time  $4.5\mu$sec 
passed through the heater, where 
the  pulse shape was 
determined by  the relaxation time $2\mu$sec of the 
current amplifier.   The total heat supplied 
was $Q=129$erg.    
 The pulse directly leaving downward 
and that reflected at the top can be distinguished since 
 their peaks are separated by $2d/c$ 
(which is $6.7\mu$sec for 
$T-T_c=500$mK). Afterward,  
the two-peak pulse  thus formed moved 
with speed $c$ in the cell 
and was  gradually flattened.  In Fig.5  we 
show pulse propagation and its 
decay at $T-T_c=500$ and $35$mK 
in the cell of $1.03$cm on 
a longer time scale, where the pulses are singly-peaked 
because of the longer pulse-duration time $7\mu$sec ($>2d/c$).  
The pulse broadening 
and the tendency of homogenization become 
more enhanced on approaching the critical point. In these 
experiments,  the fluid velocity  in the pulse region 
was  of order $v=c \delta\rho/\rho \sim 10^{-2}$cm$/$sec 
and the fluid displacement was  of 
order $\Delta x =v \Delta t \sim 10^{-7}$cm 
with $\Delta t \sim 10\mu$sec  being the 
pulse-duration time.  For such extremely small 
displacements the film heater did 
not disturb the sound propagation 
 \cite{mechanical}. 
We are interested in  the adiabatically 
increased energy $E_{\rm ad}(=p \Delta x)$ in 
 the pulse region per unit area. To linear order 
it is expressed as     
\be 
E_{\rm ad}=\frac{p}{\rho}\int dx \delta\rho(x,t) 
 = \frac{pc}{\rho}\int dt \delta\rho(x,t) .
\en 
The ratio of $E_{\rm ad}$ to the total heat 
supplied $Q=\int dt {\dot Q}(t)$ represents 
the efficiency of transforming applied  heat to mechanical 
work. In our experiments 
 $E_{\rm ad}/Q$ was about $0.11-0.12$  at the first 
arrival of  pulses at the cell center 
over  wide ranges of $T-T_c$ and  $Q$. However, 
a pulse has a tail persisting in  later times 
and there remains ambiguity of order $10\%$ 
in the calculation of  $E_{\rm ad}$.

\begin{figure}[h]
\includegraphics[scale=0.5]{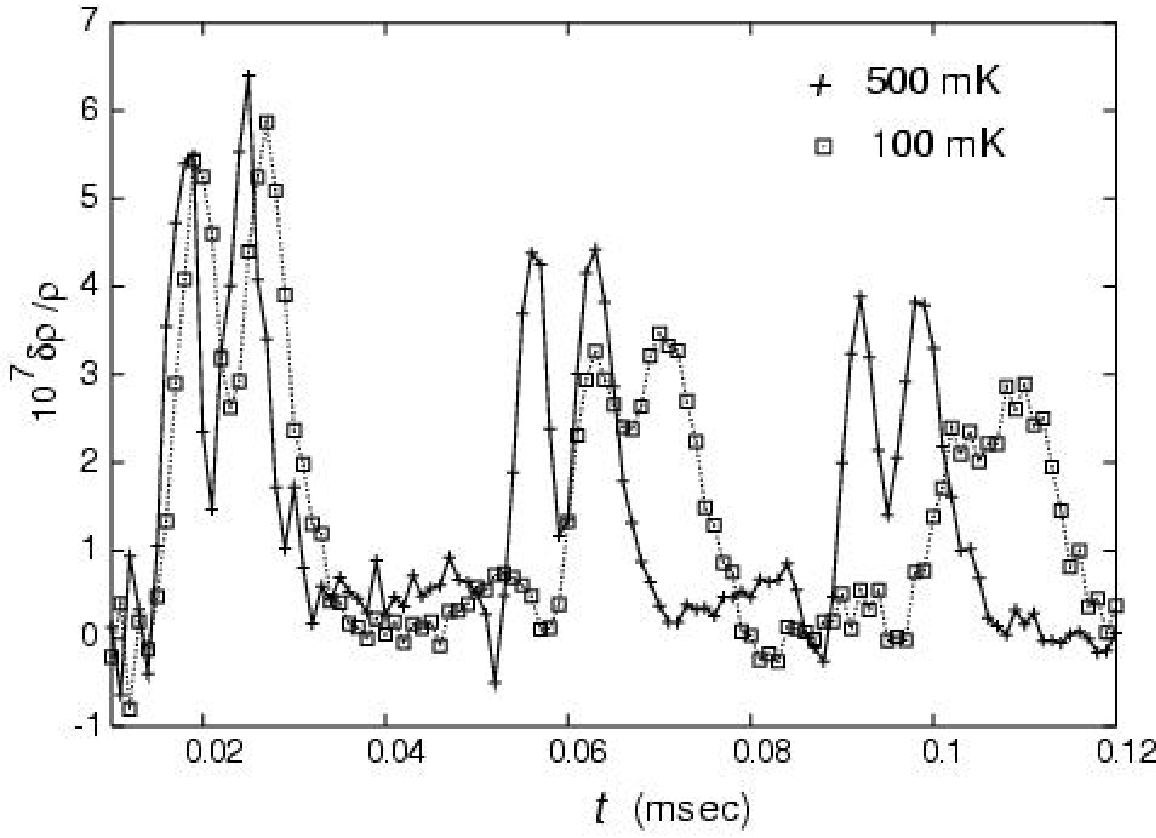}
\caption{
 $\delta\rho(t)/\rho$ 
at the cell center for 
$T-T_c=500$mK 
and $100$mK in the time region 
$0<t<0.12$msec,  produced by pulse heating of 
width $4.5\mu$sec in a cell of $L=5.5$mm.}
\end{figure}

\begin{figure}[h]
\includegraphics[scale=0.5]{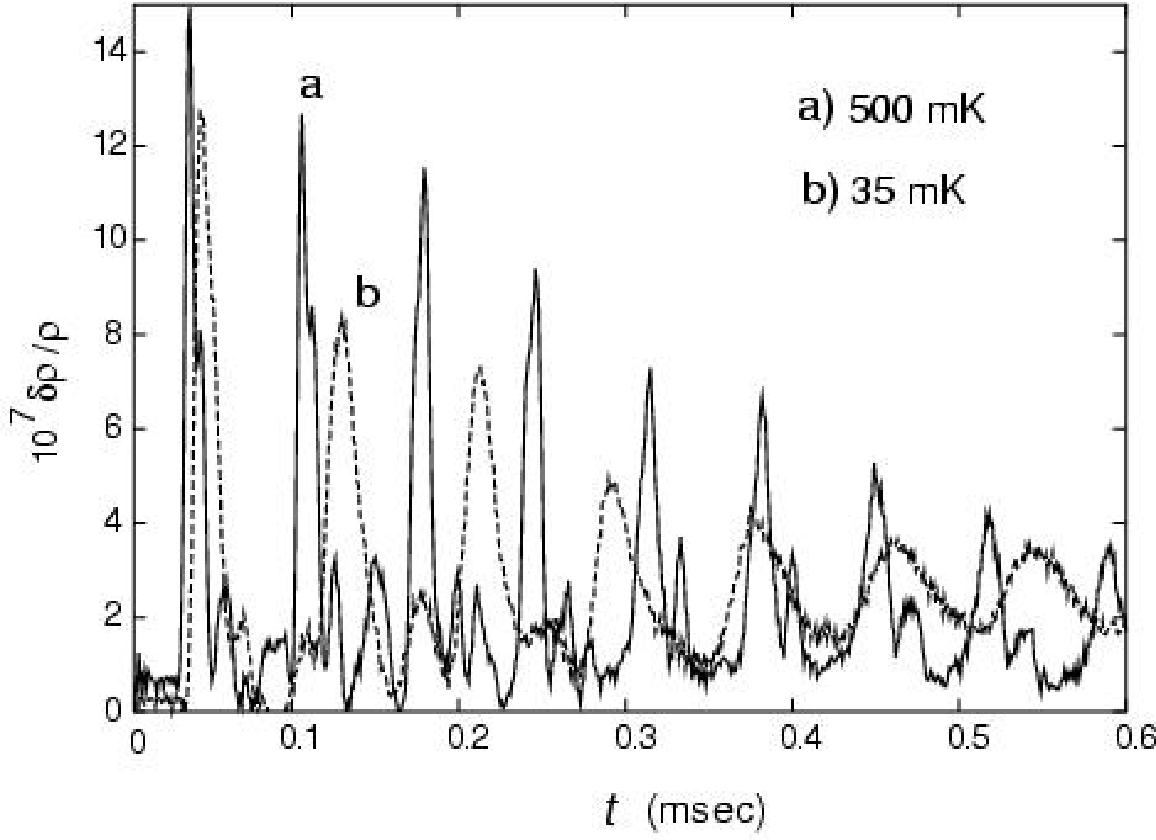}
\caption{
$\delta\rho(t)/\rho$ 
at the cell center for 
$T-T_c=500$mK 
and $35$mK in the time region $0<t<0.6$msec,  
produced by pulse heating of 
width $7\mu$sec in a cell of $L=1.03$cm.}
\end{figure}

We give theoretical interpretations of the above results 
 neglecting the effect of the bulk viscosity. 
Let a small amount of heat be supplied to 
 a near-critical fluid 
at a rate ${\dot Q}(t)$ per unit area for $t>0$ 
in a one-dimensional geometry. 
The volume expansion rate per unit area near the heater is 
\be 
{\dot V}_1= 
\ppp{ \rho^{-1}}{s}{p} \frac{{\dot Q}(t)}{ T}, 
\en 
where  $s$ is the entropy per 
unit mass. If the heater is attached to 
the upper plate, this volume change 
then produces a sound wave propagating 
downward as $\delta\rho_{\rm out}(t-x/c)$, 
where $x$ is the distance from the heater 
and $\delta\rho_{\rm out}(t)$ is the density 
increase in the sound emitted 
at the heater at time $t$. 
Since acoustic disturbances 
move with the sound speed $c$, 
the mass conservation 
gives the sound amplitude,   
\be 
\delta \rho_{\rm out}(t)=\frac{\rho}{c}{\dot V}_1
= \frac{\rho}{cT}\ppp{T}{p}{s} \dot{Q}(t),  
\en 
where  the Maxwell relation 
$(\p \rho^{-1}/\p s)_p= (\p T/\p p)_s$ has been used. 
The adiabatic coefficient 
$(\p T/\p p)_s$ is nearly equal to 
the derivative along the coexistence curve  
 $(\p T/\p p)_{\rm cx}$  
near the critical point \cite{Onukibook}. 
For   CO$_2$  it is equal to $
T_c/6.98p_c$\cite{Hohenberg} 
and  the above relation 
becomes 
$\delta \rho_{\rm out}/\rho = 1.38\times 10^{-13}\dot{Q}$ 
with $\dot Q$ in erg$/$cm$^2$sec. 
The heating rate 
used for our data in Fig.2 was 
  $\dot{Q}=0.183\times 10^7$  and  then 
$\delta \rho_{\rm out}/\rho = 2.55\times 10^{-7}$.  
This theoretical value fairly agrees with 
the observed height of the first step 
$\sim 2.2\times 10^{-7}$ in Fig.2.  
The efficiency of energy transformation 
discussed below Eq.(2) becomes \cite{Ferrell} 
\be 
\frac{E_{\rm ad}}{Q} = \frac{p}{T} \ppp{T}{p}{s}, 
\en 
which is $1/6.98= 0.14$  for 
near-critical CO$_2$.  
On the other hand, 
the  experimental values were 
in the range $0.11-0.12$.

As a complicating factor in our experiment, 
however, the heater is placed $d=0.5$mm below 
the top plate. If ${\dot V}_2$  
is the volume expansion rate 
in the thermal diffusion layer  at the top plate, 
we require 
$
c\delta \rho_{\rm out}(t)/\rho ={\dot V}_1 +{\dot V}_2$. 
In the time region where 
$(Dt)^{1/2} \ll d \ll ct < L$, 
some calculations yield a modified convolution-type formula, 
\be 
\delta \rho_{\rm out}(t)
= \frac{\rho}{cT}\ppp{T}{p}{s}
\int_0^t dt'  \dot{Q}(t-t')\dot{\psi}(t'/t_2)/t_2,  
\en 
where $t_2= L^2/4c^2t_1$ 
is a  characteristic time of heat 
exchange between the boundary and  
sounds  in the range  $L/c \ll t_2\ll t_1$.  
The scaling  function 
$\dot{\psi}(u)$ is positive 
and its integral $\psi(u)= \int_0^udv \dot{\psi}(v)$
tends to 1 for large $u$. The function  
$\psi(u)$ itself appeared in the original theory 
 \cite{Ferrell}. 
Thus Eq.(5) reduces to  
Eq.(3) when ${\dot Q}(t)$ changes much slower than 
$t_2$.

We clarify   the relationship 
of the formula (3) and the assumption in the 
theory of the piston effect \cite{Ferrell}.
Let the heat input 
$\dot{Q}(t)$ changes slowly compared with 
the acoustic time $t_a=L/c$. 
We suppose  a time interval with width 
$\delta t \gg t_a$, in which  $\dot{Q}(t)$ is almost 
unchanged.  After many sound traversals,   
the adiabatic pressure and  density 
increases in the interior region are given by 
\be 
\delta p= c^2 \delta\rho= 
c^2\frac{\delta t}{t_a} 
\frac{\rho}{cT}\ppp{T}{p}{s} \dot{Q}, 
\en 
as a result of superposition of many steps. 
In terms of 
the incremental 
heat supply $\delta Q= \dot{Q}\delta t$
  we find 
\be 
\delta p 
=\frac{\rho }{LT}\ppp{T}{\rho}{s} \delta{Q} 
= \ppp{p}{s}{\rho}\frac{\delta Q}{\rho TL},
\en 
where  the Maxwell relation 
 $(\p p/\p s)_\rho= \rho^2 
(\p T/\p \rho)_s$  has been used. 
If $\delta p$ is assumed to be 
homogeneous, Eq.(8) follows 
from the space average of 
$
\delta p= 
(\p{p}/\p{s})_{\rho}\delta s+ 
(\p{p}/\p{\rho})_{s}\delta \rho  
$ \cite{Ferrell},   
since the space integral of $\delta s$ is 
$\delta Q/\rho T$  and that of $\delta\rho$ 
vanishes.  In addition, the average temperature deviation 
is  written as  $\av{\delta T}= \delta Q/C_VL$ 
where $C_V=\rho T (\p s/\p T)_\rho$ 
is the isochroic specific heat \cite{Onukibook}.  
From Eq.(3) we may thus reproduce  the 
original theory of the piston effect.

In summary, we have first 
measured rapid acoustic responses 
in near-critical CO$_2$ to extremely small heat inputs, 
which constitute a basis of understanding 
adiabatic processes in fluids.  We demonstrate that 
adiabatic changes take place  on the timescale of $L/c$. 
As Fig.2 suggests,  it is intriguing 
that  a homogeneous 
adiabatic change can be achieved  almost instantaneously 
 if the  heating time is  a multiple of 
$2L/c$.  The theoretical expressions (3) and (4) are 
useful to analyze 
 these results and can reproduce 
the original theory of the piston effect valid on long timescales. 
The theoretical 
 height of a stepwise sound in Eq.(4) 
and  efficiency $E_{\rm ad}/Q$ in Eq.(5) 
 fairly  agree with  
the experimental values. 
Differences of $20-30\%$ 
remain,    
probably  because  a fraction of  heat 
escaped to the stainless part 
and  the pulse shape was 
considerably  broadened even at the first arrival at 
the cell center.

There are a number of unsolved problems. 
(i) We should understand how a sound pulse 
is reflected at a boundary wall and 
how it is damped in the bulk region. 
In our case, the change of the pulse shape on reflection 
is apparently more marked than that due to the bulk damping, as 
can be  inferred from comparison of 
 the results for $L=5.5$mm and those 
for $L=1.03$cm.    This is still 
the case even very close to  the critical point. 
(ii) The bulk viscosity  exhibits 
strong critical anomaly as 
$\zeta \cong  
0.03\rho c^2 t_\xi$ \cite{Onukibook}, 
where $t_\xi$ is  the  
order parameter relaxation time 
($18\mu$sec at $T-T_c=30$mK) 
\cite{Swinney}.  It gives  a bulk acoustic damping, 
but it also  affects the 
thickness of the thermal diffusion layer 
when the typical timescale is 
longer than $\gamma \zeta/\rho c^2 
\cong 0.03\gamma t_\xi$ \cite{Carles,Moldover}. 
The effects of the bulk viscosity 
on heat transport are  not yet well investigated, however.   
(iii) 
Upon heat exchange 
 the wall temperature 
is not fixed  
when the effusivity ratio  
$a_w=  (\lambda_wC_w/\lambda C_p)^{1/2}$ 
is small \cite{Hao,Moldover}. Here
$\lambda_w$,  $\lambda$, 
$C_w$, and $C_p$ are  the  thermal 
conductivities  and the specific 
heats (per unit volume) of  the solid and fluid,
 respectively.   For finite $a_w$ the piston time 
 is  given by    
${t_1}'= [(1+a_w^{-1})/(\gamma-1)]^2L^2/4{D}$ \cite{Hao},  
which becomes  $t_1'\cong C_V^2 L^2/4\lambda_wC_w$ 
for $a_w\ll 1$. In our case 
 $a_w=3\times 10^3(T/T_c-1)^{0.92}$ between Cu and CO$_2$ 
and $a_w <1$ is reached for $T-T_c < 50$mK. 
This crossover should be studied.

 This work was  
 was initiated as a research project 
 of the National Space Development Agency of Japan 
and was afterward supported by grants from the Japan Space Forum 
   and   the Ministry of Education, 
Culture, Sports, Science and Technology of Japan.
Thanks are also due to Takeo Satoh  for 
valuable discussions.



\begin{references}
\bibitem{Ferrell} 
A. Onuki, H. Hao, and R. A. Ferrell, Phys. Rev. A {\bf 41}, 
2256 (1990); A. Onuki and R.A. Ferrell,
Physica A {\bf 164}, 245 (1990). 
If the boundary temperature is changed by $T_1$, 
the interior temperature deviation 
relaxes as $T_1\psi(t/t_1)$,  
where $\psi(u)$ is a scaling  function 
behaving as $2(u/\pi)^{1/2}$ for $u\ll 1$ 
and  $1- (\pi u)^{-1/2}$ for $u\gg 1$.
\bibitem{Ch} We compare the time derivative 
of the heat stored in the heater 
$C_h \p T/\p t$ and 
the heat flux $\lambda \p T/\p x$ at  the heater. 
If $\omega$ is the typical frequency, 
the ratio of the former to the latter is  
$ 2\times 10^{-2} (T/T_c-1)^{1.24}
(\omega/D)^{1/2}\ll 1$.
\end{references}
\end{document}